\documentclass[twocolumn]{aastex631}
\graphicspath{{./}{figures/}}
\usepackage{mathtools}
\usepackage{amsmath}

\begin{document}
\title{\huge Mixing Paint: \\\Large An analysis of color value transformations in multiple coordinate spaces using multivariate linear regression}

\author{Alexander Messick}
\affiliation{Washington State University, Department of Physics and Astronomy, WA 99164, USA}

\begin{abstract}
I explore the mathematical transformation that occurs in color coordinate space when physically mixing paints of two different colors. I tested 120 pairs of 16 paint colors and used a linear regression to find the most accurate combination of input parameters, both in RGB space and several other color spaces. I found that the fit with the strongest coefficient of determination was a geometrically symmetrized linear combination of the colors in CIEXYZ space, while this same mapping in RGB space returns a better mean squared error.
\end{abstract}

\section{Introduction} \label{sec:intro}
From childhood, we are taught a simple model of colors: red plus yellow makes orange, yellow and blue make green, and so on.
However, the scientific theory of colors has been found to be more complicated \citep{Opticks}.
Namely, we know that light is merely the visible portion of the electromagnetic spectrum and that different wavelengths of light are absorbed by specialized receptors in the eye and interpreted as color \citep{doi:10.1098/rstl.1802.0004}.
This model of light is known as additive color and is well understood by even premodern physics \citep{1853AnP...165...69G, maxwell_1857}.

One example of the additive color model is the modern RGB system common in electronic screens.
Each color in this system is described as a combination of red, green, and blue light, each able to take a value from 0 to 255 in intensity.
This simple model can differentiate $2^{24}\sim1.7\times10^7$ colors, each taking no more than 3 bytes of storage.

Contrary to the additive framework, there exists what is called subtractive color mixing.
In this model, colors arise from the combinations of pigments which absorb certain wavelengths of light and reflect others.
Thus, colors are not determined by the mixture of light but the lack thereof.
This is further affected by various real world factors that range from mundane to quite complicated.
For instance, paints with identical RGB values can mix with other paints to produce entirely different colors due to the chemical and physical properties of the paints themselves.

\begin{figure}
    \centering
    \includegraphics[width=\columnwidth]{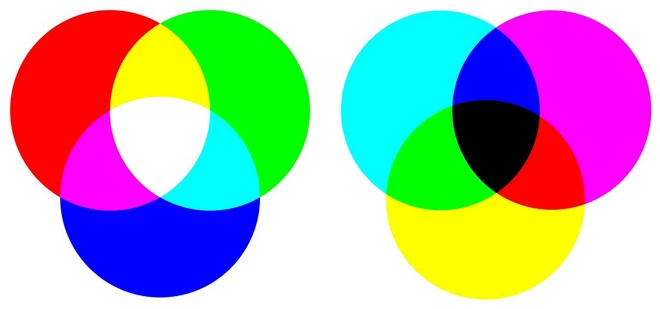}
    \caption{A comparison of additive (RGB) and subtractive (CMY) color spaces. The overlap of two circles indicate the combinations of the colors involved. In these cases, the primary colors of one model are the secondary colors of another. Image taken from \cite{Canon}.}
    \label{fig:rgb}
\end{figure}

An example of subtractive colors is the CMY model.
In this system, colors are represented as mixtures of pigments, whose ``primary'' colors are cyan, magenta and yellow.
Interestingly, these three colors correspond to the secondary colors of the RGB system, as shown in Figure \ref{fig:rgb}.
Technically, the combination of the first three pigments should absorb all visible light, showing black (called ``key''), which can be included as its own pigment.
This is known as the CYMK model, which I will examine in depth in Section \ref{subsec:convert}, along with other systems.

This paper is organized as follows: in Section \ref{sec:proc}, I discuss how the paints are mixed and applied to the canvas.
I also give the mathematical conversions from the RGB color space to various others.
In Section \ref{sec:analysis}, I explain how the resulting colors are read in as RGB coordinates and what kind of mathematical tests I will apply to the data.
In Sections \ref{sec:results} and \ref{sec:discuss}, I give the results of the analyses and determine what is the most accurate predictor of color mixing.

\subsection{Mathematical Notation}\label{subsec:math_not}
Note that throughout this paper, I use vector and matrix arithmetic in line with how arrays and matrices function in the NumPy package \citep{harris2020array} of Python (and most coding languages), where operations are performed element-wise on a data structure.
That is, for the vector/array $\vec{v} \coloneqq (v_0,v_1,v_2)$, then
\begin{equation}
    \begin{array}{c}
        \vec{v}^2 \coloneqq (v_0^2,v_1^2,v_2^2) \\
        \log{\vec{v}} \coloneqq (\log{v_0},\log{v_1},\log{v_2}) \\
        \end{array},
\end{equation}
where $\log$ refers to the natural logarithm (base $e$).
I will also use the broadcasting rules of NumPy, which are summarized as follows:
\begin{itemize}
    \item Operations between objects of the exact same shape are performed element-wise
    \item If one object can be ``stretched'' (duplicated) along one or more higher axes to match the shape of another, operations can be performed (element-wise with respect to the ``stretched'' object)
\end{itemize}
Here, the term ``higher'' refers to the order of the axes, beginning with the rightmost index in coding structures.
For examples of broadcasting:
\begin{equation}
    \vec{v} + 1 \coloneqq (v_0,v_1,v_2) + (1,1,1) = (v_0+1,v_1+1,v_2+1)
\end{equation}
For the matrix
\begin{equation}
    \begin{array}{c}
        M=
        \begin{pmatrix}
            M_{00} & M_{01} & M_{02} \\
            M_{10} & M_{11} & M_{12} \\
            M_{20} & M_{21} & M_{22} \\
        \end{pmatrix} \\
        M + \vec{v} =
        \begin{pmatrix}
            M_{00} & M_{01} & M_{02} \\
            M_{10} & M_{11} & M_{12} \\
            M_{20} & M_{21} & M_{22} \\
        \end{pmatrix} +
        \begin{pmatrix}
            v_0 & v_1 & v_2 \\
            v_0 & v_1 & v_2 \\
            v_0 & v_1 & v_2 \\
        \end{pmatrix} \\
        =
        \begin{pmatrix}
            M_{00}+v_0 & M_{01}+v_1 & M_{02}+v_2 \\
            M_{10}+v_0 & M_{11}+v_1 & M_{12}+v_2 \\
            M_{20}+v_0 & M_{21}+v_1 & M_{22}+v_2 \\
        \end{pmatrix}
    \end{array}
\end{equation}
As such, the multiplication of appropriate matrices occurs element-wise, rather than a traditional matrix multiplication.
For this reason, traditional matrix multiplications are represented with the `@' operator (apologies to the annoyed mathematicians).
\begin{equation}
    \begin{array}{c}
        M \vec{v} =
        \begin{pmatrix}
            M_{00}v_0 & M_{01}v_1 & M_{02}v_2 \\
            M_{10}v_0 & M_{11}v_1 & M_{12}v_2 \\
            M_{20}v_0 & M_{21}v_1 & M_{22}v_2 \\
        \end{pmatrix} \\
        M @ \,\vec{v} =
        \begin{pmatrix}
            M_{00} & M_{01} & M_{02} \\
            M_{10} & M_{11} & M_{12} \\
            M_{20} & M_{21} & M_{22} \\
        \end{pmatrix}
        \begin{pmatrix}
            v_0 \\
            v_1 \\
            v_2 \\
        \end{pmatrix} \\
        =
        \begin{pmatrix}
            M_{00}v_0 + M_{01}v_1 + M_{02}v_2 \\
            M_{10}v_0 + M_{11}v_1 + M_{12}v_2 \\
            M_{20}v_0 + M_{21}v_1 + M_{22}v_2 \\
        \end{pmatrix}
        
    \end{array}
\end{equation}

\section{Procedure}\label{sec:proc}
\subsection{Painting}\label{subsec:paint}
I begin with sixteen unique colors of acrylic paint, all of which were produced by Apple Barrel craft paints.
I used the same brand of paint for all colors so that physical properties such as density, viscosity, water content, etc. were relatively uniform.
The colors are listed in Table \ref{tab:paints} alongside their respective expected RGB values according to \cite{encycolor}, an online catalog containing data pertaining to various paints and colors.

\begin{deluxetable}{l c c c c}[htb]
\tablecolumns{5}
\tablecaption{Paints Used in Analysis \label{tab:paints}}
\tablehead{\colhead{Name} & \colhead{Code} & \colhead{(r,} & \colhead{g,} & \colhead{$b)_{enc}$}}
\startdata
    Fuchsia          & 20216E & (251, &  93, & 177) \\
    Purple Pansy     & 21488  & (64,  &  20, & 103) \\
    Admiral Blue     & 21484  & (35,  &  62, & 153) \\
    Too Blue         & 20771  & (9,   &  40, & 104) \\
    Tropical Blue    & 20348  & -     & -    & -    \\
    Holly Branch     & 21478  & (1,   &  96, & 91)  \\
    Lime Tree        & 21476  & (176, & 194, & 54)  \\
    Yellow Flame     & 21474  & -     & -    & -    \\
    King's Gold      & 20760  & (254, & 195, & 57)  \\
    Jack-o-lantern   & 21472  & (247, & 148, & 30)  \\
    Flag Red         & 21469  & (183, &  44, & 38)  \\
    Barn Red         & 20577  & (128, &   0, & 15)  \\
    Melted Chocolate & 20258  & (85,  &  59, & 35)  \\
    Black            & 20504  & (0,   &   0, & 0)   \\
    Pavement         & 21490  & (55,  &  50, & 78)  \\
    White            & 20503  & (255, & 255, & 255)
\enddata
\tablecomments{A table giving the name, product code, and RGB values reported by encycolorpedia for each of the paints used in this study.}
\end{deluxetable}

Next, I divide an 8" by 10" canvas panel into 256 equal rectangles (measuring 1/2" by 5/8" each).
Given 16 colors, there are 120 different combinations of paint that can be made with a simple 1:1 ratio (excluding self-combinations).
Each of these derived colors is created by adding one drop from the bottle of each parent paint to a palette and mixing until homogeneous.
Unfortunately, this introduces variety into the ratio of paints used in each combination as it is unlikely that each drop containing the same volume.

Regardless, each paint and combination thereof are systematically applied to the canvas.
I paint the diagonal cells of the canvas with the sixteen original paints, and every off-diagonal element is a mix of the two corresponding diagonals.
For example, if admiral blue is in the third diagonal cell and white is in the last, then the cell corresponding to last element of third row is painted with a roughly equal mixture of admiral blue and white.
To avoid redundancy and to leave room for further experimentation, only the upper right portion of the canvas is painted in this phase, thus the column index will be greater than or equal to the row number.
Each cell receives at least two coats of its respective paint with some receiving more based on visual inspection.

\begin{figure}
   \centering
   \includegraphics[width=\columnwidth]{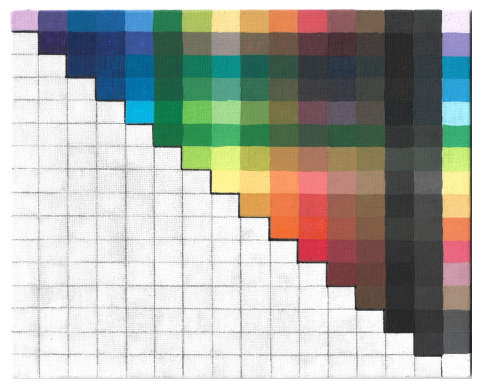}
   \caption{The final version of the painted canvas used for analysis.}
   \label{fig:canvas}
\end{figure}

At this point, I perform a rudimentary analysis of the colors present, the details of which are given in Section \ref{subsec:deter}.
From this analysis, I am able to determine a standard deviation of the RGB values for the pixels in each cell, giving an objective standard of homogeneity.
I chose to paint an additional layer for each cell for which $||\sigma_{RGB}||\gtrsim15$.
The final results of the painting process are shown in Figure \ref{fig:canvas}.

\subsection{Conversions from RGB}\label{subsec:convert}
Along with the RGB system, there are additional color spaces worth considering, like the previously mentioned CMY(K) model.
Unfortunately, many of these conversions can be equipment dependent, both in recording and creating colors, so I will use na\"ive conversions when possible.
Many of these conversions were taken from \cite{rapidtables}.
For mathematical purposes, it is convenient to write the RGB values (as with other color values) as a vector,
\begin{equation}
    \vec{c}_{RGB} \coloneqq (R,G,B).
\end{equation}
Similarly, I can define a vector for the uncertainties:
\begin{equation}
    \vec{\sigma}_{RGB} \coloneqq (\sigma_R,\sigma_G,\sigma_B)
\end{equation}

\subsubsection{CMY and CMYK}

The na\"ive conversion from RGB to CMY can be done easily with the following formula:
\begin{equation}
    \vec{c}_{CMY} \coloneqq 255 - \vec{c}_{RGB}
\end{equation}
As such, the uncertainties follow simply from
\begin{equation}
    \sigma_{f}^2 = \sum_{i=R,G,B} \left(\frac{\partial f}{\partial i}\sigma_{i}\right)^2,
\end{equation}
\begin{equation}
    \vec{\sigma}_{CMY} = \vec{\sigma}_{RGB}
\end{equation}
Note that sometimes, these values are normalized to unity rather than 255, but this does not affect the analysis, so I chose to disregard this.

A conversion to CMYK is dependent on which color has the highest value.
The ``key'' value is the complement of the highest value normalized to one.
The remaining colors are then defined and re-normalized by this maximum value $m=\max(\vec{c}_{RGB})$.
\begin{equation}
    \vec{c}_{CMYK} \coloneqq 1 - 
    \begin{pmatrix}
        \frac{\vec{c}_{RGB}}{m} \\
        \frac{m}{255}
    \end{pmatrix}
    =
    \begin{pmatrix}
        1 - \frac{R}{m} \\
        1 - \frac{G}{m} \\
        1 - \frac{B}{m} \\
        1 - \frac{m}{255}
    \end{pmatrix}
\end{equation}
By definition, at least one of the resulting values for C, M, or Y must be zero, since any nonzero combination of all three would ideally result in black.
Accordingly, the uncertainties are given by:
\begin{equation}
    \begin{array}{c}
        \vec{\sigma}_{CMYK}^2 = 
        \begin{pmatrix}
            \left(\frac{\vec{\sigma}_{RGB}}{m}\right)^2 + \left(\frac{\vec{c}_{RGB}}{m^2}\sigma_m\right)^2 \\
            \left(\sigma_m/255\right)^2
        \end{pmatrix} \\
        =
        \begin{pmatrix}
            \left(\frac{\sigma_R}{m}\right)^2 + \left(\frac{R}{m^2}\sigma_m\right)^2 \\
            \left(\frac{\sigma_G}{m}\right)^2 + \left(\frac{G}{m^2}\sigma_m\right)^2 \\
            \left(\frac{\sigma_B}{m}\right)^2 + \left(\frac{B}{m^2}\sigma_m\right)^2 \\
            \left(\sigma_m/255\right)^2
        \end{pmatrix} \\
    \end{array}  
\end{equation}.

\subsubsection{Hue-Saturation-Value (and Projection)}
\begin{figure}
    \centering
    \includegraphics[width=\columnwidth]{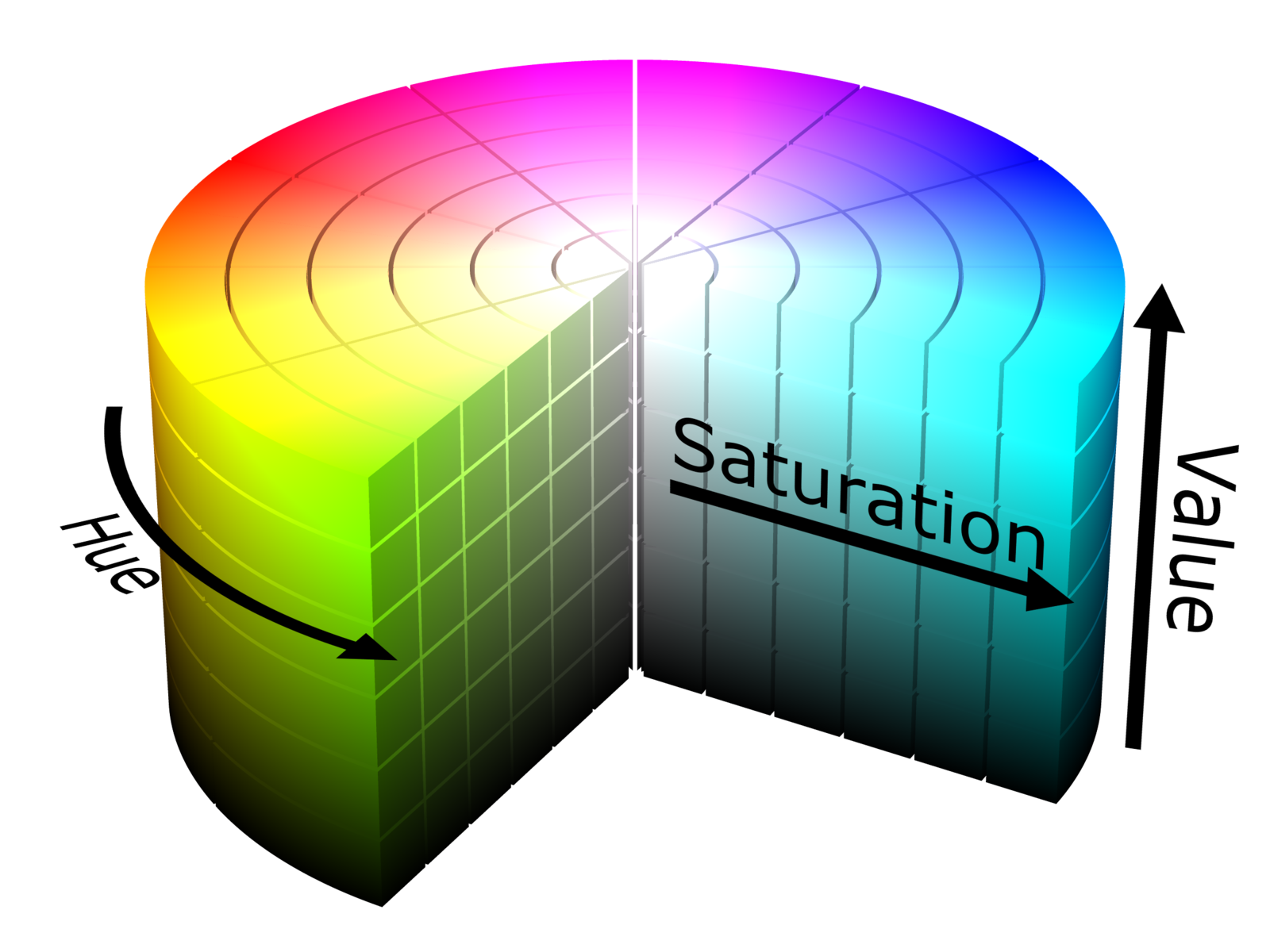}
    \caption{The HSV color wheel, with axes labeled. Note that red, green, and blue are equidistant, as opposed to a traditional color wheel where red, yellow, and blue are equidistant. Image taken from \cite{HSV_wheel}.}
    \label{fig:hsv}
\end{figure}

The Hue-Saturation-Value model for colors is a cylindrical model, similar to a familiar color wheel.
Hue is the azimuthal angle with red, green, and blue at 0, $\frac{2\pi}{3}$, and $\frac{4\pi}{3}$ radians, respectively\footnote{For a typical color wheel, red, yellow, and blue are equidistant, as they are the ``primary'' colors.}.
The radial distance represent saturation, or the color's ``intensity''.
Finally, the height represents the value, or ``brightness''.
This is shown in Figure \ref{fig:hsv}.
Mathematically, this conversion is done as follows (with $m=\max(\vec{c}_{RGB})$ once again):
\begin{equation}
    \Delta \coloneqq \max(\vec{c}_{RGB}) - \min(\vec{c}_{RGB})
\end{equation}
\begin{equation}
    H \coloneqq
    \begin{cases}
        0, & \Delta = 0 \\
        \frac{\pi}{3}\frac{G-B}{\Delta}, & m = R \\
        \frac{\pi}{3}\left(\frac{B-R}{\Delta}+2\right), & m = G \\
        \frac{\pi}{3}\left(\frac{R-G}{\Delta}+4\right), & m = B
    \end{cases}
\end{equation}
\begin{equation}
    S \coloneqq
    \begin{cases}
        0, & m = 0 \\
        \frac{\Delta}{m}, & m > 0
    \end{cases}
\end{equation}
\begin{equation}
    V \coloneqq \frac{m}{255}
\end{equation}
Typically, the angle H is taken modulo $2\pi$ so that it falls in the range $[0,2\pi)$, but this only complicates further calculations, so I will ignore it here.
Unfortunately, the if-then nature of these definitions makes the calculus of uncertainty propagation particularly atrocious here.
The values for $\Delta$ and H depend on which colors have the highest and lowest values, making derivation tricky.
The different cases can be summarized with a few matrices.
First, I define a base matrix depending on which color is the maximum value (assuming $m\neq0$ and $\Delta\neq0$):
\begin{equation}
    M_{HSV} \coloneqq
    \begin{cases}
        \begin{pmatrix}
            0 & \frac{\pi}{3S} & -\frac{\pi}{3S} \\
            -S & 0 & 0 \\
            \frac{m}{255} & 0 & 0
        \end{pmatrix}, & m = R \\
        \begin{pmatrix}
             -\frac{\pi}{3S} & 0 & \frac{\pi}{3S} \\
            0 & -S & 0 \\
            0 & \frac{m}{255} & 0
        \end{pmatrix}, & m = G \\
        \begin{pmatrix}
            \frac{\pi}{3S} & -\frac{\pi}{3S} & 0\\
             0 & 0 & -S \\
             0 & 0 & \frac{m}{255}
        \end{pmatrix}, & m = B
    \end{cases}
\end{equation}

From there, the columns vector $\vec{v}$ is added to the column corresponding to the maximum color and subtracted to that of the minimum color, then the entire matrix is divided by the maximum value.

\begin{equation}
    \vec{v}_{HSV} \coloneqq
    \begin{cases}
        \begin{pmatrix}
            -\frac{H}{S} \\ 1 \\ 0
        \end{pmatrix}, & m = R \\
        \begin{pmatrix}
            \frac{1}{S} (\frac{2\pi}{3} - H) \\ 1 \\ 0
        \end{pmatrix}, & m = G \\
        \begin{pmatrix}
            \frac{1}{S} (\frac{4\pi}{3} - H) \\ 1 \\ 0
        \end{pmatrix}, & m = B
    \end{cases}
\end{equation}

For example, for the case $R<G<B\neq0$:
\begin{equation}
    \begin{array}{c}
        M_{HSV+} \coloneqq \frac{1}{m} \Biggl[
        \begin{pmatrix}
            \frac{\pi}{3S} & -\frac{\pi}{3S} & 0\\
             0 & 0 & -S \\
             0 & 0 & \frac{m}{255}
        \end{pmatrix} \\
        +
        \begin{pmatrix}
            -\frac{1}{S} (\frac{4\pi}{3} - H) & 0 & \frac{1}{S} (\frac{4\pi}{3} - H) \\
             -1 & 0 & 1 \\
             0 & 0 & 0
        \end{pmatrix} \Biggr] \\
        = \frac{1}{m}
        \begin{pmatrix}
            \frac{H-\pi}{S} & -\frac{\pi}{3S} & \frac{1}{S} (\frac{4\pi}{3} - H) \\
            -1 & 0 & 1-S \\
            0 & 0 & \frac{m}{255}
        \end{pmatrix} \\
        =
        \begin{pmatrix}
            \frac{\pi}{3}\frac{B-G}{(B-R)^2} & -\frac{\pi}{3}\frac{1}{B-R} & -\frac{\pi}{3}\frac{R-G}{(B-R)^2} \\
            -\frac{1}{B} & 0 & \frac{R}{B^2} \\
            0 & 0 & \frac{1}{255}
        \end{pmatrix}
    \end{array}
\end{equation}
Finally, this matrix is squared element-wise and right multiplied by the RGB uncertainties to give the HSV uncertainties:
\begin{equation}
    \sigma_{HSV}^2 = 
    \begin{pmatrix}
        \sigma_H^2 \\
        \sigma_S^2 \\
        \sigma_V^2
    \end{pmatrix}
    = M_{HSV+}^2 @\, \sigma_{RGB}^2
\end{equation}

Typically, when color mixing is done on a color wheel, it is by averaging the positions of the initial colors on the wheel, a task which is easier done in Cartesian coordinates.
From this cylindrical system to Cartesian, we can simply use the following:
\begin{equation}
    \vec{c}_{XYV} =
    \begin{pmatrix}
        X \\
        Y \\
        V
    \end{pmatrix} = 
    \begin{pmatrix}
        S\cos{H} \\
        S\sin{H} \\
        V
    \end{pmatrix}
\end{equation}
where `Y' represents the Y-position, not to be confused with the yellow value in the CMY(K) system.
Just as with the HSV uncertainties, the uncertainties of these new coordinates are mathematically arduous.
There are nine cases, but they can be summarized with a few matrices, similar to the previous conversion but with more steps.
First,
\begin{equation}
    \Lambda \coloneqq
    \begin{cases}
        \begin{pmatrix}
            -S & 0 & 0 \\
            0 & \frac{\pi}{3} & -\frac{\pi}{3} \\
            \frac{m}{255} & 0 & 0
        \end{pmatrix}, & m = R \\
        \begin{pmatrix}
            0 & -S & 0 \\
            -\frac{\pi}{3} & 0 & \frac{\pi}{3} \\
            0 & \frac{m}{255} & 0
        \end{pmatrix}, & m = G \\
        \begin{pmatrix}
             0 & 0 & -S\\
            \frac{\pi}{3} & -\frac{\pi}{3} & 0 \\
            0 & 0 & \frac{m}{255}
        \end{pmatrix}, & m = B \\
    \end{cases}
\end{equation}
Then the vector $\vec{v}_{XYV}$ is added to the column corresponding to the maximum value and subtracted from column corresponding to the minimum value, where

\begin{equation}
    \vec{v}_{XYV} \coloneqq
    \begin{cases}
    
        \begin{pmatrix}
            1 \\
            -H \\
            0
        \end{pmatrix}, & m = R \\
        \begin{pmatrix}
             1 \\
            \frac{2\pi}{3} - H  \\
            0
        \end{pmatrix}, & m = G \\
        \begin{pmatrix}
            1 \\
            \frac{4\pi}{3} - H \\
            0
        \end{pmatrix}, & m = B
    \end{cases}
\end{equation}

For example, for the case $R < G < B \neq 0$ :
\begin{equation}
    \begin{array}{c}
        \Lambda_+ = \frac{1}{m} \left[
        \begin{pmatrix}
             0 & 0 & -S\\
            \frac{\pi}{3} & -\frac{\pi}{3} & 0 \\
            0 & 0 & \frac{m}{255}
        \end{pmatrix} +
        \begin{pmatrix}
            -1 & 0 &  1 \\
             H-\frac{4\pi}{3} & 0 & \frac{4\pi}{3}-H \\
             0 & 0 & 0
        \end{pmatrix} \right]\\
        = \frac{1}{m}
        \begin{pmatrix}
            -1 & 0 & 1-S \\
            H-\pi & -\frac{\pi}{3} & \frac{4\pi}{3}-H \\
            0 & 0 & \frac{m}{255}
        \end{pmatrix} \\
        = \frac{1}{B}
        \begin{pmatrix}
            -1 & 0 & \frac{R}{B} \\
            -\frac{\pi}{3}\frac{G-B}{B-R} & -\frac{\pi}{3} & -\frac{\pi}{3}\frac{R-G}{B-R} \\
            0 & 0 & \frac{B}{255}
        \end{pmatrix}
    \end{array}
\end{equation}

From here, the resulting matrix is left multiplied by the matrix $R_Z(H)$ corresponding to a rotation an angle H about the z-axis, later divided by the maximum color value:
\begin{equation}
    M_{XYV+} \coloneqq R_Z(H) @ \, \Lambda_+,
\end{equation}
where
\begin{equation}
    R_Z(H) \coloneqq
    \begin{pmatrix}
        \cos{H} & -\sin{H} & 0\\
        \sin{H} & \cos{H} & 0 \\
        0 & 0 & 1
    \end{pmatrix}.
\end{equation}
Finally, the matrix is squared element-wise and matrix multiplied by a vector containing RGB variances to produce the X and Y variances.

\begin{equation}
    \vec{\sigma}_{XYV}^2=
    \begin{pmatrix}
        \sigma_X^2 \\
        \sigma_Y^2 \\
        \sigma_V^2
    \end{pmatrix} = 
    M_{XYV+}^2 @ \, \vec{\sigma}_{RGB}^2
\end{equation}

This all seems very complicated, but luckily it is simple logic to code, significantly easier than performing the calculus on each case separately and coding in the results (I checked).

\subsubsection{CIE Color Spaces}
The CIELAB (also known as L*a*b*) system and its parent CIEXYZ are color spaces defined by the International Commission on Illumination (CIE).
They are intended to be perceptually uniform color spaces, meaning a change in numerical space yields a similarly perceived change in color.
Unlike other color systems, the conversion from RGB to the CIELAB system is not dependent on equipment, but the ``standard observer'' (more or less convention).

First, the RGB values must be converted to CIEXYZ.
This can be done by first gamma-expanding (a process called ``linearizing'') each color, then multiplying by a matrix:
\begin{equation}
    C_{lin} =
    \begin{cases}
        \frac{C}{3294.6} & C \leq 10.31475 \\
        \left( \frac{C+14.025}{269.025}\right)^{2.4} & C > 10.31475
    \end{cases}
\end{equation}
\begin{equation}
    \begin{array}{c}
         \vec{c}_{XYZ} = M_{65} @\, \vec{c}_{RGB,lin} =\\
         \begin{pmatrix}
            X_{65} \\ Y_{65} \\ Z_{65}
        \end{pmatrix} =
        \begin{pmatrix}
            0.4124 & 0.3576 & 0.1805 \\
            0.2126 & 0.7152 & 0.0722 \\
            0.0193 & 0.1192 & 0.9505
        \end{pmatrix}
        \begin{pmatrix}
            R_{lin} \\ G_{lin} \\ B_{lin}
        \end{pmatrix}
    \end{array}
\end{equation}
$M_{65}$ is the matrix for the BT.709 primaries defined by the Radiocommunication sector of the International Telecommunication \cite{BT.709}.
The `65' refers to the CIE standard illuminant D65, an illumination convention roughly corresponding to average midday light in Northwestern Europe.
The uncertainties are fairly straight forward:
\begin{equation}
    C'_{lin} =
    \begin{cases}
        \frac{1}{3294.6} & C \leq 10.31475 \\
        \frac{24}{2690.25}C_{lin}^{7/12} & C > 10.31475
    \end{cases}
\end{equation}
\begin{equation}
    \vec{\sigma}_{XYZ}^2 = [M_{65}\vec{c}_{RGB,lin}^\prime]^2 @\, \vec{\sigma}_{RGB}^2 \\
\end{equation}

The CIELAB space converts these coordinates onto axes corresponding to lightness and two color gradients (green to red and blue to yellow).
\begin{equation}
    L^* = 166 f\left(\frac{Y}{100}\right) - 16
\end{equation}
\begin{equation}
    a^* = 500\left[f\left(\frac{X}{95.0489}\right) - f\left(\frac{Y}{100}\right) \right]
\end{equation}
\begin{equation}
    b^* = 200\left[f\left(\frac{Y}{100}\right) - f\left(\frac{Z}{108.884}\right) \right]
\end{equation}
where
\begin{equation}
    f(t) =
    \begin{cases}
        \frac{t}{3\delta^2}+\frac{4}{29}, & t\leq\delta^3 \\
        \sqrt[3]{t}, & t>\delta^3
    \end{cases}
\end{equation}
and $\delta=\frac{6}{29}$.
The uncertainties are then:
\begin{equation}
    \sigma_L^2 = \left[\frac{166}{100} g\left(\frac{Y}{100}\right)\sigma_Y\right]^2
\end{equation}
\begin{equation}
    \sigma_a^2 = \left[\frac{500}{95.0489} g\left(\frac{X}{95.0489}\right)\sigma_X\right]^2 +  \left[5g\left(\frac{Y}{100}\right)\sigma_Y\right]^2
\end{equation}
\begin{equation}
    \sigma_b^2 = \left[2g\left(\frac{Y}{100}\right)\sigma_Y\right]^2 +  \left[\frac{200}{108.884}g\left(\frac{Z}{108.884}\right)\sigma_Z\right]^2
\end{equation}
where
\begin{equation}
    g(t) = f^\prime(t) =
    \begin{cases}
        \frac{1}{3\delta^2}, & t\leq\delta^3 \\
        \frac{1}{3}t^{-2/3}=\frac{f(t)}{3t}, & t>\delta^3
    \end{cases}
\end{equation}

\section{Analysis}\label{sec:analysis}
\subsection{Determining Colors}\label{subsec:deter}
I scanned the canvas using an HP DeskJet 2652 \citep{DeskJet} and saved the data as a PNG file.
The image was pre-processed (rotated and cropped to be 1600 by 2000 pixels), and divided into its 256 cells, each 100 by 125 pixels.
To account for border spillover (i.e., painting outside of the lines), each cell was given a pixel buffer along its perimeter.
I tested various buffer sizes, and for each, I calculated the magnitude of the standard deviation of the RGB values in each cell and found the maximum.
I then selected buffer size to minimize this maximum standard deviation, resulting in a size of 20 pixels.

\begin{figure}
    \centering
    \includegraphics[width=\columnwidth]{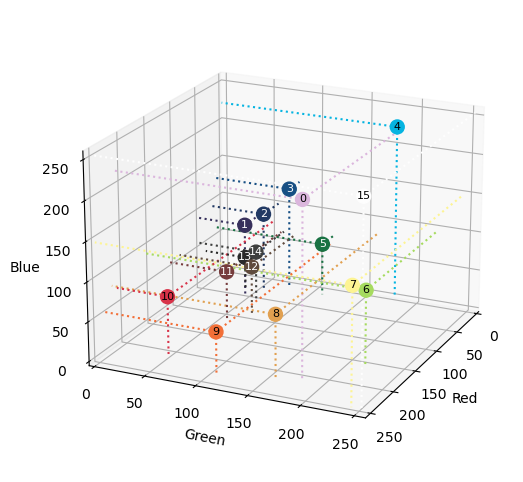}
    \caption{A 3-D plot of the resulting RGB values for the sixteen base colors.}
    \label{fig:plot}
\end{figure}

\begin{figure*}
    \centering
    \textbf{Measured and Na\"ively Predicted RGB Values}
    \includegraphics[width=\textwidth]{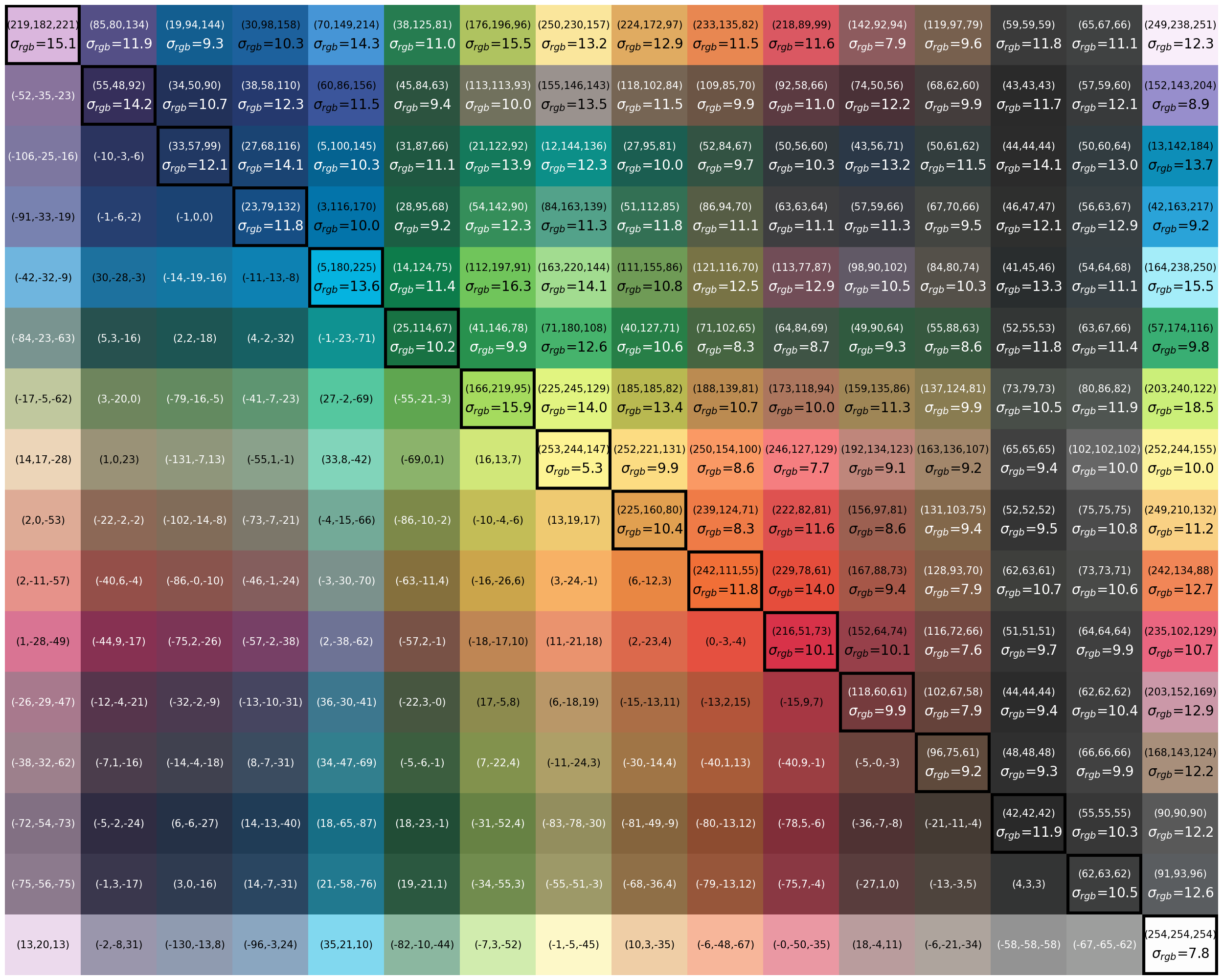}
    \caption{An image showing the color coordinates for all of the colors analyzed. Cells along the diagonal (encased in black borders) depict the sixteen base colors, and off-diagonal colors show the mix of the corresponding diagonal colors. The upper right triangle (including the diagonal) shows the final measured RGB values for the resulting color and the magnitude of the standard deviations thereof. Cells in the lower left are colored according to the expected RGB values (assuming a simple mean) and show the difference between the measured and predicted values.}
    \label{fig:grid}
\end{figure*}

Having applied this buffer, I analyzed the remaining pixels inside each cell.
The mean and standard deviation of the RGB values are calculated and recorded as vectors.
After this initial analysis, I repainted any cells for which the magnitude of this standard deviation was above 15, as mentioned in Section \ref{sec:proc}.
I then re-scanned and processed the image, repeating the same color analysis.
Even after this repainting, there still remained a handful of cells for which $||\sigma_{RGB}||\gtrsim15$.
The final RGB coordinates for the base colors are plotted in Figure \ref{fig:plot}, and the RGB values and standard deviation magnitudes are shown in Figure \ref{fig:grid}.

\subsection{Mathematical Exploration}\label{subsec:math}
I explore a few possibilities for the transformation from two input colors, represented numerically, to their corresponding output. 

First, I test if the output color is a linear combination of their inputs, that is:
\begin{equation}
    \begin{array}{l}
        R_3=\alpha_{1R} R_1 + \beta_{1R} G_1 + \gamma_{1R} B_1 + \alpha_{2R} R_2 + \cdots \\
        G_3=\alpha_{1G} R_1 + \beta_{1G} G_1 + \gamma_{1G} B_1 + \alpha_{2G} R_2 + \cdots \\
        B_3=\alpha_{1B} R_1 + \beta_{1B} G_1 + \gamma_{1B} B_1 + \alpha_{2B} R_2 + \cdots
        \end{array}
\end{equation}
or more succinctly:
\begin{equation}
    \vec{c}_3 = \Theta_1 @ \,\vec{c}_1 + \Theta_2 @ \,\vec{c}_2
\end{equation}
where $\vec{c}_i = (R_i,G_i,B_i)$ and
\begin{equation}
    \Theta_i =
    \begin{pmatrix}
        \alpha_{iR} & \beta_{iR} & \gamma_{iR} \\
        \alpha_{iG} & \beta_{iG} & \gamma_{iG} \\
        \alpha_{iB} & \beta_{iB} & \gamma_{iB} \\
    \end{pmatrix}.
\end{equation}

The other main framework I explore is that the resulting color is a geometric combination of the inputs.
\begin{equation}
    \begin{array}{l}
        R_3= R_1^{\alpha_{1R}} G_1^{\beta_{1R}} B_1^{\gamma_{1R}} \times R_2^{\alpha_{2R}} G_2^{\beta_{2R}} B_2^{\gamma_{2R}} \\
        G_3= R_1^{\alpha_{1G}} G_1^{\beta_{1G}} B_1^{\gamma_{1G}} \times R_2^{\alpha_{2G}} G_2^{\beta_{2G}} B_2^{\gamma_{2G}} \\
        B_3= R_1^{\alpha_{1B}} G_1^{\beta_{1B}} B_1^{\gamma_{1B}} \times R_2^{\alpha_{2B}} G_2^{\beta_{2B}} B_2^{\gamma_{2B}}
        \end{array}
\end{equation}

There is no convenient matrix equation in this form, but I can take the logarithm of both sides to get:

\begin{equation}
    \begin{array}{l}
        \log{R_3}=\alpha_{1R} \log{R_1} + \beta_{1R} \log{G_1} + \gamma_{1R} \log{B_1} + \cdots \\
        \log{G_3}=\alpha_{1G} \log{R_1} + \beta_{1G} \log{G_1} + \gamma_{1G} \log{B_1} + \cdots \\
        \log{B_3}=\alpha_{1B} \log{R_1} + \beta_{1B} \log{G_1} + \gamma_{1B} \log{B_1} + \cdots
        \end{array}
\end{equation}
or 
\begin{equation}
    \log{\vec{c}_3} = \Theta_1 @ \,\log{\vec{c}_1} + \Theta_2 @ \,\log{\vec{c}_2}
\end{equation}

An unfortunate side effect of this notation, however, is that it can only handle positive values.
This means that it does not work without modification for zeros.
To avoid this problem, I introduce an arbitrarily small number $\epsilon=10^{-16}$ to replace every instance of zero prior to testing.
This does not, however, solve the problem for negative values (such as the XYV and CIELAB systems), so these cannot be used with geometric fitting.

It is worth noting that in the RGB system, both the arithmetic and geometric means of mathematically combining colors have agreements and disagreements with traditional intuition.
For instance, if one were to take the arithmetic mean of the two input colors, red (255,0,0) and blue (0,0,255) make a medium-dark purple (128,0,128) as one might expect, but red and cyan (0,255,255), green (0,255,0) and magenta (255,0,255), and blue and yellow (255,255,0) make the same dull grey (128,128,128).
Taking the geometric mean yields even stranger results: any combination of the primary colors or any combination involving black (0,0,0) can only produce black.

Additionally, there are a few conditions I choose to impose:
\begin{enumerate}
    \item Symmetry: $\vec{c}_3=f(\vec{c}_1,\vec{c}_2)=f(\vec{c}_2,\vec{c}_1)$
    \item Completeness: $\sum_{i=1}^{6}a_{ij}=1$
\end{enumerate}

In other words, there is no physical reason that the order of inputting paints should affect the outcome; red and blue should produce the same color as blue and red.
Mathematically, this symmetry can be imposed several ways, so I choose to test linear or geometric symmetrization:
\begin{equation}
    \vec{c}_{sym,lin} \coloneqq \frac{\vec{c}_1 + \vec{c}_2}{2}
\end{equation}
or
\begin{equation}
    \vec{c}_{sym,geo} \coloneqq \sqrt{\vec{c}_1\vec{c}_2}
\end{equation}
so now
\begin{equation}
    \vec{c}_{3} \coloneqq \Theta @\, \vec{c}_{sym}
\end{equation}
Just as with the case with geometric fitting, geometric symmetrization breaks down for negative values (since they can result in imaginary components), and so it cannot be used with the XYV and CIELAB systems.

The second condition imposes that the sum for the coefficients in any aforementioned equation is unity.
This way, the mathematical range [0,255] remains the same before and after transformation, for both the arithmetic and geometric cases.
This condition is less physically motivated, so I examine the results with and without completeness imposed.
Mathematically, this can be imposed with $\gamma_i=1-\alpha_i-\beta_i$:
\begin{equation}
    \vec{c}_3 =
    \begin{pmatrix}
        \alpha_{R} & \beta_{R} & 1 - \alpha_{R} - \beta_{R} \\
        \alpha_{G} & \beta_{G} & 1 - \alpha_{G} - \beta_{G} \\
        \alpha_{B} & \beta_{B} & 1 - \alpha_{B} - \beta_{B} \\
    \end{pmatrix}
    \begin{pmatrix}
        R \\ G \\ B
    \end{pmatrix}_{sym}
\end{equation}
which can be rewritten as 
\begin{equation}
    \vec{c}_3 - B_{sym}=
    \begin{pmatrix}
        \alpha_{R} & \beta_{R} \\
        \alpha_{G} & \beta_{G} \\
        \alpha_{B} & \beta_{B} \\
    \end{pmatrix}
    \begin{pmatrix}
        R - B \\ G -B 
    \end{pmatrix}_{sym}
\end{equation}
Though the equations above are notated with $R,G,B$ values, this same mathematical framework can be used for any of the color systems we analyze.
For the CMYK case, $\Theta$ becomes a $4\times4$ matrix, but all other calculations remain the same.
Nevertheless, I use the LinearRegression function from scikit-learn to fit each system I wish to analyze.
This is a ordinary least-squares regression that returns a coefficient matrix and prediction score.
In order to get a distribution of scores, we repeatedly sample the data from a normal distribution centered around the reported value and whose standard deviation is equal to the reported uncertainty.

\section{Results} \label{sec:results}
I use coefficients of determination (r-score) as measure for the wellness of a fit.
The results for all of the different tests can be seen in Table \ref{tab:score}.

\begin{deluxetable*}{c | C || CCCCCCCC }[htb]
    \rotate
    \tablecolumns{10}
    \tablecaption{Analysis Methods and Scores \label{tab:score}}
    \tablehead{
        \colhead{}      & \colhead{Method:}   & \colhead{Linear} & \colhead{Linear} &\colhead{Geom.}  & \colhead{Geom.}  & \colhead{Linear} & \colhead{Linear} & \colhead{Geom.}  & \colhead{Geom.} \\
        \colhead{Color} & \colhead{Symmetry:} & \colhead{Arith.} & \colhead{Geom.}  & \colhead{Arith.} & \colhead{Geom.}  & \colhead{Arith.} & \colhead{Geom.}  & \colhead{Arith.} & \colhead{Geom.} \\
        \colhead{}      & \colhead{Complete:} & \colhead{Comp.}  & \colhead{Comp.}  & \colhead{Comp.}  & \colhead{Comp.}  & \colhead{Free}   & \colhead{Free}   & \colhead{Free}   & \colhead{Free} \\
        (1) & & (2) & (3) & (4) & (5) & (6) & (7) & (8) & (9)
        }
    \startdata
        R  && 0.727 \pm 0.012 & 0.789 \pm 0.015 & 0.564 \pm 0.034 & 0.547 \pm 0.042 & 0.714 \pm 0.009 & \mathbf{0.797 \pm 0.013} & 0.506 \pm 0.037 & 0.492 \pm 0.047 \\
        G  && 0.732 \pm 0.019 & 0.758 \pm 0.023 & 0.676 \pm 0.026 & 0.696 \pm 0.030 & 0.835 \pm 0.010 & \mathbf{0.886 \pm 0.009} & 0.788 \pm 0.017 & 0.845 \pm 0.015 \\
        B  && 0.122 \pm 0.019 & 0.090 \pm 0.023 & 0.206 \pm 0.026 & 0.144 \pm 0.030 & 0.649 \pm 0.017 & \mathbf{0.748 \pm 0.018} & 0.631 \pm 0.021 & 0.703 \pm 0.022 \\
        \hline
        C  && 0.727 \pm 0.012 & 0.506 \pm 0.026 & 0.738 \pm 0.082 & 0.278 \pm 0.104 & 0.714 \pm 0.009 & 0.451 \pm 0.018 & 0.779 \pm 0.079 & 0.438 \pm 0.074 \\
        M  && 0.732 \pm 0.020 & 0.540 \pm 0.041 & 0.701 \pm 0.077 & 0.174 \pm 0.105 & 0.835 \pm 0.010 & 0.663 \pm 0.021 & 0.835 \pm 0.039 & 0.488 \pm 0.074 \\
        Y  && 0.122 \pm 0.019 & 0.253 \pm 0.030 & 0.044 \pm 0.029 & 0.083 \pm 0.073 & 0.649 \pm 0.017 & 0.548 \pm 0.022 & 0.694 \pm 0.061 & 0.389 \pm 0.102 \\
        \hline
        C && 0.629 \pm 0.023 & 0.030 \pm 0.023 & 0.046 \pm 0.049 & 0.055 \pm 0.046 & 0.525 \pm 0.028 & 0.018 \pm 0.013 & 0.053 \pm 0.049 & 0.059 \pm 0.047 \\
        M && 0.703 \pm 0.024 & 0.034 \pm 0.027 & 0.047 \pm 0.042 & 0.054 \pm 0.043 & 0.418 \pm 0.042 & 0.058 \pm 0.021 & 0.051 \pm 0.042 & 0.059 \pm 0.044 \\
        Y && 0.825 \pm 0.018 & 0.032 \pm 0.024 & 0.055 \pm 0.051 & 0.059 \pm 0.047 & 0.629 \pm 0.036 & 0.206 \pm 0.031 & 0.061 \pm 0.053 & 0.063 \pm 0.047 \\
        K && 0.008 \pm 0.005 & 0.029 \pm 0.022 & 0.007 \pm 0.007 & 0.019 \pm 0.017 & 0.757 \pm 0.011 & 0.556 \pm 0.023 & 0.669 \pm 0.137 & 0.317 \pm 0.302 \\
        \hline
        H  && 0.068 \pm 0.085 & 0.041 \pm 0.072 & 0.096 \pm 0.117 & 0.065 \pm 0.084 & 0.103 \pm 0.086 & 0.066 \pm 0.073 & 0.135 \pm 0.114 & 0.104 \pm 0.087 \\
        S  && 0.243 \pm 0.033 & 0.017 \pm 0.016 & 0.035 \pm 0.050 & 0.032 \pm 0.046 & 0.346 \pm 0.037 & 0.248 \pm 0.028 & 0.050 \pm 0.055 & 0.048 \pm 0.050 \\
        V  && 0.104 \pm 0.024 & 0.018 \pm 0.020 & 0.025 \pm 0.026 & 0.026 \pm 0.024 & 0.761 \pm 0.013 & 0.791 \pm 0.012 & 0.698 \pm 0.019 & 0.774 \pm 0.016 \\
        \hline
        X  && 0.610 \pm 0.028 &                 &                 &                 & 0.633 \pm 0.029 &                 &                 & \\
        Y  && 0.541 \pm 0.045 &                 &                 &                 & 0.501 \pm 0.048 &                 &                 & \\
        V  && 0.013 \pm 0.007 &                 &                 &                 & 0.743 \pm 0.011 &                 &                 & \\
        \hline
        X  && 0.701 \pm 0.016 & 0.804 \pm 0.020 & 0.573 \pm 0.019 & 0.690 \pm 0.021 & 0.761 \pm 0.007 & \mathbf{0.897 \pm 0.007} & 0.701 \pm 0.011 & 0.853 \pm 0.009 \\
        Y  && 0.665 \pm 0.019 & 0.762 \pm 0.028 & 0.586 \pm 0.022 & 0.669 \pm 0.026 & 0.763 \pm 0.010 & \mathbf{0.902 \pm 0.009} & 0.732 \pm 0.014 & 0.863 \pm 0.010 \\
        Z  && 0.157 \pm 0.023 & 0.050 \pm 0.029 & 0.228 \pm 0.024 & 0.069 \pm 0.019 & 0.601 \pm 0.015 & \mathbf{0.791 \pm 0.017} & 0.641 \pm 0.018 & 0.743 \pm 0.016 \\
        \hline
        $L^*$ && 0.799 \pm 0.020 &              &                 &                 & 0.763 \pm 0.010 &                 &                 & \\
        $a^*$ && 0.617 \pm 0.038 &              &                 &                 & 0.347 \pm 0.046 &                 &                 & \\
        $b^*$ && 0.197 \pm 0.035 &              &                 &                 & 0.477 \pm 0.022 &                 &                 & \\
    \enddata
    \tablecomments{A table giving the scores for the different mathematical methods for paint combination. The Col. (1) gives the color, horizontally separated by color space. The remaining rows in the table header give the method of analysis (linear vs. geometric combination), the symmetrization used (arithmetic or geometric), and the completeness imposed (complete or free). All values are given with $1-\sigma$ uncertainties with the two best performing fits (discussed in Sections \ref{sec:results} and \ref{sec:discuss}) have been made bold.}
\end{deluxetable*}

The color set that had the best overall scores, both in mean and magnitude, corresponds to a geometrically symmetrized linear combination (GSLC) of the CIEXYZ color space.
In other words:
\begin{equation}
    \vec{c}_{3,XYZ} = \Theta_{XYZ} @\, \sqrt{\vec{c}_{1,XYZ}\vec{c}_{2,XYZ}}
\end{equation}
where $\Theta_{XYZ} =$
\begin{equation}\label{xyz_coeff}
    \begin{pmatrix}
         0.882 \pm 0.078 &  0.095 \pm 0.085 & 0.044 \pm 0.026 \\
        -0.058 \pm 0.089 &  1.021 \pm 0.100 & 0.098 \pm 0.037 \\
         0.211 \pm 0.096 & -0.307 \pm 0.105 & 1.037 \pm 0.042
    \end{pmatrix}
\end{equation}
Analyzing the rows of this matrix, we see mathematically that the largest contributor to each output ``color'' is the input value itself (X vs. X, etc.).
With the exception of the bottom row, all off-diagonal elements have a magnitude less than 0.1 and all diagonal elements but the first are within 0.1 of unity.
The resulting Z-coordinate, however, is more dependent on the other values, as evidenced by the bottom row of the matrix.
Because coefficients of this fit were free and not complete, the output values can exceed the accepted range for the system and will need to be dealt with (usually with a cap).

\section{Discussion}\label{sec:discuss}
Similar to the best overall result, the best fit (by mean score) that was achieved without converting from RGB space is a GSLC of the input colors:
\begin{equation}
    \vec{c}_{3,RGB} = \Theta_{RGB} @\, \sqrt{\vec{c}_{1,RGB}\vec{c}_{2,RGB}}
\end{equation}
where $\Theta_{RGB} =$
\begin{equation}\label{rgb_coeff}
    \begin{pmatrix}
        0.812 \pm 0.023 &  0.292 \pm 0.042 & 0.077 \pm 0.041 \\
        0.065 \pm 0.014 &  0.917 \pm 0.029 & 0.108 \pm 0.031 \\
        0.096 \pm 0.015 & -0.075 \pm 0.031 & 0.955 \pm 0.032
    \end{pmatrix}
\end{equation}
This scheme has a mean square error (MSE) of 2208.1 compared to the best overall fit with 0.0158.
However, scaling the RGB values to unity yields a much improved MSE of 0.0340.
Interestingly, if one were to use the results of the overall best fit and convert the predicted mixed colors back to RGB space, one would arrive at an MSE of 0.0407.
Because this MSE is actually smaller, then for the purposes of predicting RGB values, a geometrically symmetrized linear combination of the RGB values is a better predictor than the so-called ``best-fit''.

\begin{figure*}
    \centering
    \textbf{Best Fit Predicted Colors}
    \includegraphics[width=\textwidth]{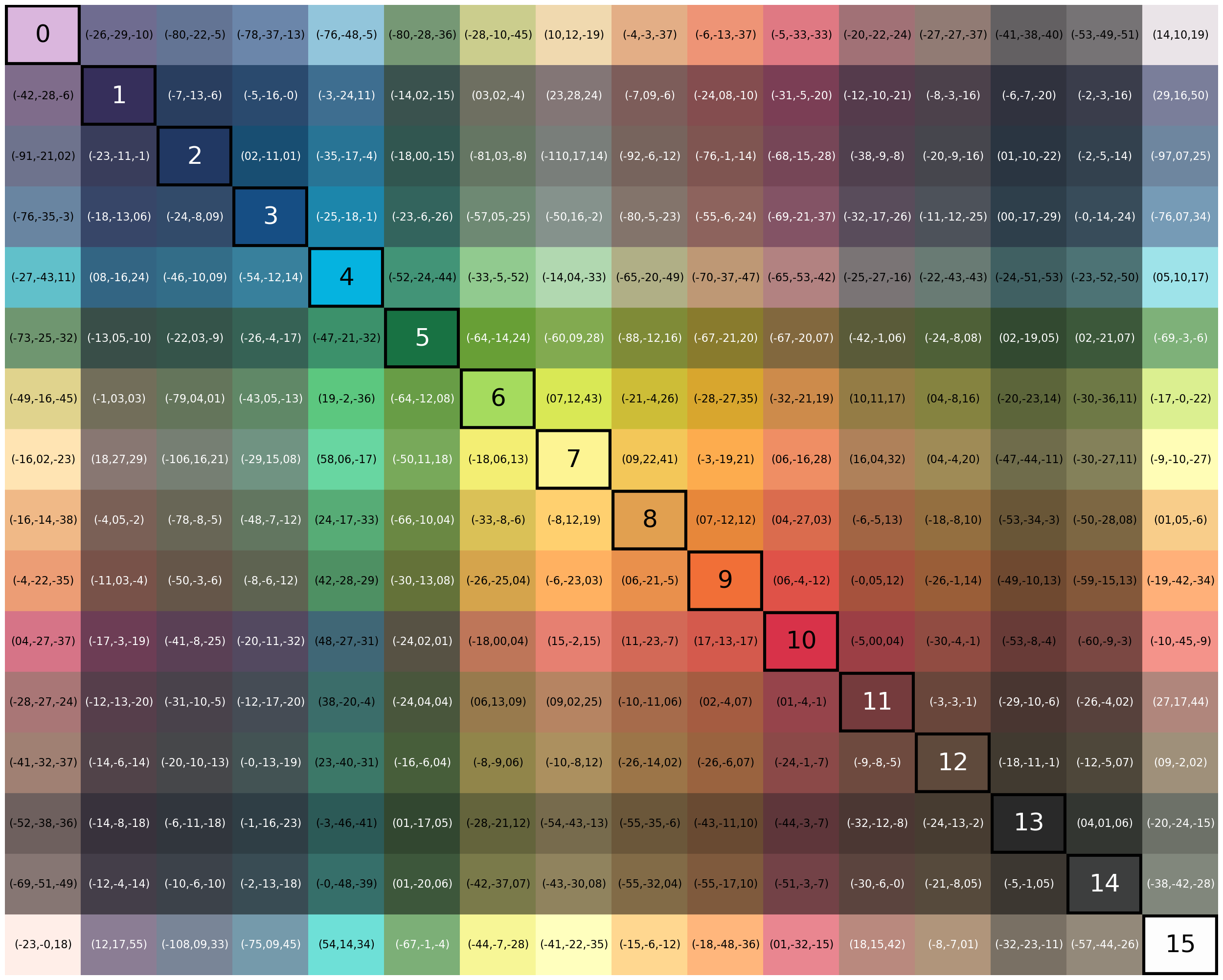}
    \caption{An image showing the predicted outcome of the different color combinations. Once again, the diagonal cells represent the sixteen base colors numbered 0 through 15, while the off-diagonal elements represent the predicted color. Cells in the upper right depict the predicted outcomes from the overall best fit (GLSC of CIEXYZ values), while cells in the lower left correspond to the best fit found without conversion (GLSC of RGB values). In both cases, we give the numerical error (the difference between the observed and predicted values).}
    \label{fig:pred}
\end{figure*}

The results of these predictions on the color mixtures can be seen in Figure \ref{fig:pred}, which can be compared with Figures \ref{fig:canvas} or \ref{fig:grid} for the real color values.
In particular, if one were to calculate the MSE for the individual colors, one would see red is consistently the most difficult value to predict.
For the GSLC of the CIEXYZ values converted back to RGB, the MSEs are 0.0253, 0.0065, and 0.0089 for red, green and blue respectively.
For the best RGB fit, the MSEs are 0.0218, 0.0058, and 0.0064.
This difference in MSE between colors would indicate that both fits had more difficulty predicting the resulting red value accurately.

Analyzing the result of each color system individually, we see fitting without conversion resulting in moderately successful predictions ($r\gtrsim0.5$) in 3 combinations, though the best was the GLSC as previously mentioned.

In the CMY space, the arithmetically symmetrized linear combination (ASLC) performed the same as the RGB values, which makes sense since they themselves are a linear combination of the RGB values (albeit with an intercept).
Both the ASLC and arithmetically symmetrized geometric combination (ASGC) of CMY values performed decently, with the ASGC fitting slightly better.
Adding the key value actually decreasing the mean score for all cases, with several individual scores below 0.1 (a very poor fit).

None of the HSV tests performed well, though this could have been expected since it is a cylindrical space which does not transform linearly or geometrically anyway.
Interestingly, all of the free tests were able to predict the V coordinate (``value'') somewhat accurately.
Because the projection of this system onto 3-D space resulted in negative values, the XYV space could only be tested as an ASLC.
This test performed well enough when the coefficients were free.

Finally, all of the tests in the CIEXYZ space with free coefficients had scores above 0.6, indicating decent to good fits.
Neither test in the CIELAB space performed particularly well.

Overall, none of the tests with complete coefficients performed well, notably for the final coordinate.
The best score for a final coordinate with complete coefficients was a GSLC of CMY values whose Y score was 0.253 which would indicate a weak correlation at best.
This problem did not occur categorically in the free-coefficient tests; only 3 of the 17 scores for the final coordinates are less than 0.5.

\section{Conclusion} \label{sec:conclusion}
I made 120 mixed colors by pairing together sixteen base paints with roughly equal proportions, all of which were applied to the same canvas panel.
This canvas was analyzed for its RGB values, which were then compared against the RGB values for each input pair.
I then tried to find the most accurate mathematical mapping from input to output colors from a small list of options.
I also converted the color values into other color spaces to test if this would improve accuracy.
In terms of mean squared error, the best performing mathematical operation to predict the outcomes of mixing two paint colors is a geometrically symmetrized linear combination of the input RGB values.
\begin{equation}
    \begin{pmatrix}
        R \\ G \\ B
    \end{pmatrix}_{out} =
    \begin{pmatrix}
        0.812 &  0.292 & 0.077 \\
        0.065 &  0.917 & 0.108 \\
        0.096 & -0.075 & 0.955
    \end{pmatrix}
    \begin{pmatrix}
        \sqrt{R_1R_2} \\ \sqrt{G_1G_2} \\ \sqrt{B_1B_2}
    \end{pmatrix}
\end{equation}
However, the strongest correlation overall was for a geometrically symmetrized linear combination of the CIEXYZ values.
\begin{equation}
    \begin{pmatrix}
        X \\ Y \\ Z
    \end{pmatrix}_{3} =
    \begin{pmatrix}
         0.882 &  0.095 & 0.044 \\
        -0.058 &  1.021 & 0.098 \\
         0.211 & -0.307 & 1.037 
    \end{pmatrix}
    \begin{pmatrix}
        \sqrt{X_1X_2} \\ \sqrt{Y_1Y_2} \\ \sqrt{Z_1Z_2}
    \end{pmatrix}
\end{equation}
In fact, the CIEXYZ demonstrated many of the strongest correlations found in this study.
Perhaps a follow-up paper examining transformations in this coordinate space could find a more accurate fit.
I would also recommend spending less time on the mathematics of coordinate transformations and their associated uncertainties and simply trusting color space conversions built into various coding packages; the Gaussian re-sampling process I mentioned in Section \ref{subsec:math} can simply be done before any conversions; there is no need to convert uncertainties.
More precision during the mixing steps could be achieved with a set of sufficiently small syringes or a sufficiently large canvas.
Finally, I would also suggest better means of photographing and recording the colors than a desktop scanner, but I received no funding for this project.

\section{Acknowledgements} \label{sec:acknowledgements}
Alexander Messick thanks the LSSTC Data Science Fellowship Program, which is funded by LSSTC, NSF Cybertraining Grant \#1829740, the Brinson Foundation, and the Moore Foundation; their participation in the program has benefited this work.
I would also like to thank my parents and my partner Heath for supporting my artistic journey, without whom I would not have previously owned the supplies used in this study and would have had to purchase them myself.

\bibliography{paint_mixer}{}
\bibliographystyle{aasjournal}

\end{document}